\documentclass{Interspeech}
\interspeechcameraready

\usepackage{cite}
\usepackage{hyperref}
\usepackage{url}
\usepackage{multirow}
\usepackage{tabularx}
\usepackage{amsmath}
\newcommand\numberthis{\addtocounter{equation}{1}\tag{\theequation}}
\DeclareMathOperator*{\argmax}{arg\,max}
\DeclareMathOperator*{\argmin}{arg\,min}
\title{Adversarial Deep Metric Learning for Cross-Modal Audio-Text Alignment in Open-Vocabulary Keyword Spotting}

\author[]{Youngmoon}{Jung}
\author[]{Yong-Hyeok}{Lee}
\author[]{Myunghun}{Jung}
\author[]{Jaeyoung}{Roh}
\author[]{Chang Woo}{Han}
\author[]{\\Hoon-Young}{Cho}
\affiliation[nocounter]{Samsung Research}{Seoul}{South Korea}
\email{\{youngm.jung, yong\_h.lee,  mh95.jung, jyo.roh, cw1105.han, h.y.cho\}@samsung.com}

\keywords{open-vocabulary keyword spotting, deep metric learning, adversarial learning, domain mismatch}

\usepackage{comment}

\begin{document}

\maketitle

\begin{abstract}
    For text enrollment-based open-vocabulary keyword spotting (KWS), acoustic and text embeddings are typically compared at either the phoneme or utterance level.
    To facilitate this, we optimize acoustic and text encoders using deep metric learning (DML), enabling direct comparison of multi-modal embeddings in a shared embedding space.
    However, the inherent heterogeneity between audio and text modalities presents a significant challenge. 
    To address this, we propose Modality Adversarial Learning (MAL), which reduces the domain gap in heterogeneous modality representations.
    Specifically, we train a modality classifier adversarially to encourage both encoders to generate modality-invariant embeddings.
    Additionally, we apply DML to achieve phoneme-level alignment between audio and text, and conduct extensive comparisons across various DML objectives.
    Experiments on the Wall Street Journal (WSJ) and LibriPhrase datasets demonstrate the effectiveness of the proposed approach.

\end{abstract}

\section{Introduction}

Keyword spotting (KWS) is the task of detecting specific keywords within audio streams and has garnered significant attention with the growing popularity of voice assistants activated by keywords such as ``Alexa,'' ``Hi Bixby,'' or ``Okay Google.'' 
KWS systems are generally classified into fixed KWS \cite{Chen14-ICASSP, Sainath-INTERSPEECH, TANG17-ICASSP}, which restricts users to a predefined set of keywords, and open-vocabulary KWS \cite{Chen-ICASSP, Huang-ICASSP, Kurmi-INTERSPEECH, Lim-arxiv, He-ICLR, Jung-INTERSPEECH, Shin-INTERSPEECH, Nishu-INTERSPEECH, Lee-INTERSPEECH, RPL-Jung-INTERSPEECH, jin24d_interspeech}, also referred to as flexible or user-defined KWS, which allows users to enroll custom keywords.
Open-vocabulary KWS can be performed via either speech-based \cite{Chen-ICASSP, Huang-ICASSP, Kurmi-INTERSPEECH, Lim-arxiv} or text-based keyword enrollment \cite{He-ICLR, Jung-INTERSPEECH, Shin-INTERSPEECH, Nishu-INTERSPEECH, Lee-INTERSPEECH, RPL-Jung-INTERSPEECH, jin24d_interspeech}.

Text-based keyword enrollment has gained popularity due to its convenience, as users can enroll keywords via text input without requiring multiple spoken examples. 
These systems typically employ a text encoder for enrollment and an acoustic encoder for inference, both of which are optimized using deep metric learning (DML) \cite{Wang-CVPR}.
Commonly used DML objectives include triplet loss \cite{He-ICLR}, Asymmetric-Proxy (AsyP) loss \cite{Jung-INTERSPEECH}, and Relational Proxy (RP) loss \cite{RPL-Jung-INTERSPEECH}.
This training process encourages acoustic embeddings (AEs) and text embeddings (TEs) of the same keyword to be closely aligned in a shared space, while embeddings of different keywords are pushed apart.
These approaches primarily focus on utterance-level embeddings (utterance-level matching), whereas another research direction explores phoneme-level audio-text alignment (phoneme-level matching).
For example, prior studies have leveraged attention mechanisms \cite{Shin-INTERSPEECH, Lee-INTERSPEECH, Ai-INTERSPEECH, Kim-SPL-2024}, dynamic programming-based algorithms \cite{Nishu-INTERSPEECH, Nishu-ICASSP}, and CTC-based alignment \cite{jin24d_interspeech} for phoneme-level matching.
Despite these advancements, limited research has investigated the integration of DML with phoneme-level matching.

Unlike traditional unimodal approaches, which rely solely on speech-based enrollment, multi-modal KWS that uses text-based enrollment faces challenges due to the inherent cross-modal heterogeneity between audio and text representations \cite{Nishu-ICASSP}.
To address this issue, we propose Adversarial Deep Metric Learning (ADML), which incorporates Modality Adversarial Learning (MAL) into the DML framework, inspired by adversarial learning \cite{Goodfellow-NIPS}.
Specifically, we train the acoustic and text encoders to confuse a modality classifier, thereby encouraging them to generate modality-invariant embeddings at both the phoneme and utterance levels, while the classifier simultaneously attempts to distinguish between audio and text modalities.
This adversarial process improves cross-modal alignment, resulting in more effective representation learning.

Our key contributions are as follows: 1) For phoneme-level matching, we apply the AsyP loss with Adaptive Margins and Scaling (AdaMS) \cite{Jung-INTERSPEECH-AdaMS} and compare its performance with various DML objectives.
2) We introduce MAL to mitigate cross-modal heterogeneity in open-vocabulary KWS.
To the best of our knowledge, this is the first work to apply adversarial learning for reducing audio-text modality mismatches in KWS. 
Additionally, we demonstrate that MAL enhances not only our method but also existing approaches, highlighting its broad applicability.
3) We incorporate a SphereFace2-based keyword classification loss \cite{Wen-ICLR}, which enhances intra-modal discrimination within the acoustic encoder. 
%using a pairwise learning strategy.
 
\begin{figure*}[t]
  \centering
  \includegraphics[width=\textwidth]{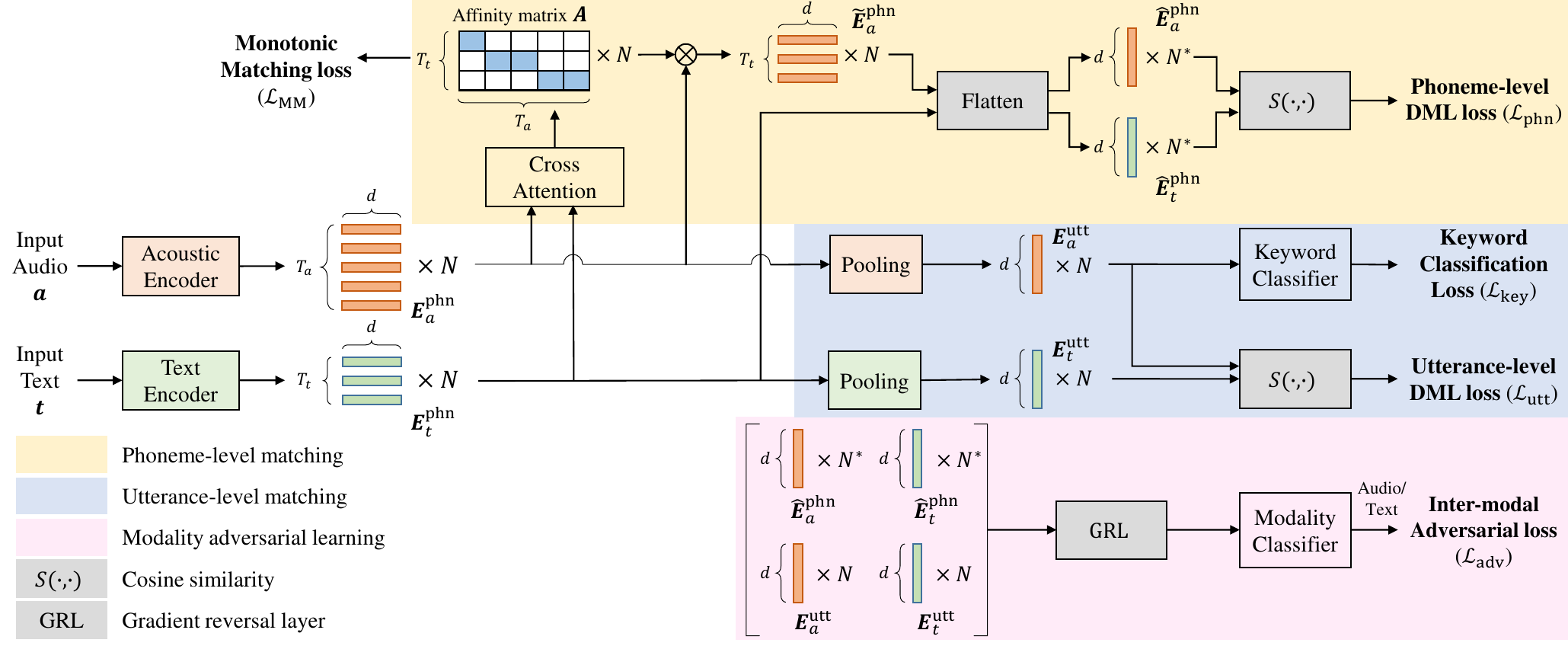}
  \vspace{-0.65cm}
  \caption{Overall architecture of Adversarial Deep Metric Learning (ADML).} 
  \label{fig:fig1}
  \vspace{-0.5cm}
\end{figure*}

\section{Related Work}

For phoneme-level matching, the Dynamic Sequence Partitioning (DSP) algorithm \cite{Nishu-INTERSPEECH, Nishu-ICASSP} aligns the acoustic sequence to match the length of the corresponding text sequence. 
After alignment, contrastive loss is applied by computing the average pairwise $L_2$ distance between aligned sequences.
Instead of using DSP, we adopt a cross-attention mechanism \cite{Shin-INTERSPEECH, Kim-SPL-2024} for alignment.
To define the phoneme-level matching loss, we use AsyP loss with AdaMS \cite{Jung-INTERSPEECH-AdaMS}, a more advanced DML approach that was originally developed for utterance-level matching.

To address cross-modal heterogeneity, Nishu \textit{et al.} \cite{Nishu-ICASSP} proposed an audio-compliant text encoder. 
Their method employs a non-trainable text encoder, where the audio encoder is pretrained using Connectionist Temporal Classification (CTC) loss. 
Phoneme-level text embeddings are then generated by averaging the embeddings of audio frames across the entire training set, mapped to their corresponding phonemes.
While effective, this approach requires a highly robust audio encoder trained on large-scale speech datasets and involves a complex process for text embedding extraction.
Additionally, it focuses exclusively on phoneme-level embeddings.
In contrast, our approach addresses cross-modal heterogeneity via MAL, which leverages adversarial training of a modality classifier that is used only during training.
This method simplifies the learning process, enhances flexibility, and enables seamless integration with various KWS methods.
Moreover, it enables the encoders to learn modality-invariant representations at both the phoneme and utterance levels, leading to improved performance.

\section{Proposed Method}

Let $\{(\bm{a}_i, \bm{t}_i, y_i) \mid i=1,\dots,N\}$ represent a batch of $N$ data samples, where $\bm{a}_i$ and $\bm{t}_i$ denote the audio and text input sequences, respectively, and $y_i$ represents the corresponding word class.
The text input $\bm{t}_i$ is a phoneme index sequence generated by a pre-trained grapheme-to-phoneme (G2P) system \cite{ParkG2P}. 
As illustrated in Fig. \ref{fig:fig1}, the audio and text inputs are processed through their respective encoders, yielding phoneme-level embeddings, $\bm{E}^\textrm{phn}_a \in \mathbb{R}^{T_a \times d}$ and $\bm{E}^\textrm{phn}_t \in \mathbb{R}^{T_t \times d}$ (for clarity, we omit the data index $i$ in the notation).
Here, $T_a$ and $T_t$ represent the sequence lengths of $\bm{E}^\textrm{phn}_a$ and $\bm{E}^\textrm{phn}_t$, respectively, and $d$, the embedding dimension, is set to 256 in this work. 
Since neither encoder contains a subsampling layer, $T_a$ and $T_t$ are equal to the lengths of their respective inputs, $\bm{a}$ and $\bm{t}$.

\subsection{Phoneme-Level Matching}

To address the sequence length disparity between $\bm{E}^\textrm{phn}_a$ and $\bm{E}^\textrm{phn}_t$, where $T_a > T_t$ in most cases, we employ a cross-attention mechanism \cite{Shin-INTERSPEECH, Kim-SPL-2024}.
In this setup, $\bm{E}^\textrm{phn}_t$ serves as the query ($Q$), while $\bm{E}^\textrm{phn}_a$ acts as both the key ($K$) and the value ($V$). 
This process computes an affinity matrix $\bm{A} \in \mathbb{R}^{T_t \times T_a}$, which aligns $\bm{E}^\textrm{phn}_a$ with $\bm{E}^\textrm{phn}_t$.
The aggregated phoneme-level acoustic embeddings $\bm{\widetilde{E}}^\textrm{phn}_a$ are then calculated as:
\begin{equation}\label{eqn:eq1}
 \bm{\widetilde{E}}^\textrm{phn}_a = \textrm{Softmax}(\frac{QK^T}{\sqrt{d_k}})\times V = \bm{A} \times V \in \mathbb{R}^{T_t \times d}.
\end{equation}
To ensure that $\bm{A}$ exhibits monotonic and diagonal alignment when $\bm{a}$ and $\bm{t}$ correspond to the same keyword (i.e., the full matching case), we apply the monotonic matching loss ($\mathcal{L}_\textrm{MM}$) proposed in \cite{Shin-INTERSPEECH}.
Unlike \cite{Shin-INTERSPEECH}, we do not use a target matrix with a random noise pattern for non-matching cases.

Using $\bm{\widetilde{E}}^\textrm{phn}_a$ and $\bm{E}^\textrm{phn}_t$, which now share the same sequence length, we define a phoneme-level DML loss ($\mathcal{L}_\textrm{phn}$). 
To facilitate this, both $\bm{\widetilde{E}}^\textrm{phn}_a$ and $\bm{E}^\textrm{phn}_t$ are flattened along their batch and sequence length dimensions using the \texttt{tf.ragged.boolean\_mask} function in TensorFlow. 
This converts their original 3D shapes ($N \times T^{\textrm{max}}_t \times d$) into 2D ragged tensors of size $N^* \times d$, where $T^{\textrm{max}}_t$ is the maximum sequence length in a batch and $N^* = \sum_{i=1}^NT_{t,i} \leq N \times T^{\textrm{max}}_t$. 
The resulting flattened embeddings, denoted as $\bm{\hat{E}}^\textrm{phn}_a$ and $\bm{\hat{E}}^\textrm{phn}_t$, correspond to individual phoneme labels. 
Similarly, $\bm{t}$ is flattened, producing $N^*$ phoneme labels, denoted as $\hat{t}$.
Thus, the mini-batch is reformulated as $N^*$ data tuples, represented as $\{(\bm{\hat{E}}_{a,i}^\textrm{phn}, \bm{\hat{E}}_{t,i}^\textrm{phn}, \hat{t}_i) \mid i=1,\dots,N^*\}$. 
The phoneme-level DML loss $\mathcal{L}_\textrm{phn}$ is defined using the AsyP loss, formulated as the sum of anchor-positive and anchor-negative terms:
\begin{align*}
    \mathcal{L}_\textrm{phn}=\frac{1}{N^*} \sum_{i=1}^{N^*} \bigg (\frac{1}{\alpha} \underset{j \in \mathcal{P}_i}{\textrm{ELSE}} \, \alpha(\lambda - S(\bm{\hat{E}}_{t,i}^\textrm{phn},\bm{\hat{E}}_{a,j}^\textrm{phn})) \\ + \underset{k \in \mathcal{N}_i}{\textrm{MSP}} \, \beta(S(\bm{\hat{E}}_{a,i}^\textrm{phn},\bm{\hat{E}}_{t,k}^\textrm{phn}) - \lambda) \bigg), \numberthis \label{eq2}
\end{align*}
where $\mathcal{P}_i = \{j \mid \hat{t}_j = \hat{t}_i\}$ and $\mathcal{N}_i = \{k \mid \hat{t}_k \neq \hat{t}_i\}$ represent the sets of positive and negative indices, respectively, and $S(\cdot, \cdot)$ denotes the cosine similarity. 
The terms $\textrm{ELSE}$ and $\textrm{MSP}$ refer to the Extended-LogSumExp function \cite{Wang-CVPR} and the Mean-Softplus function \cite{Yi-ICPR}, respectively.
The first term in $\mathcal{L}_\textrm{phn}$ draws the anchor $\bm{\hat{E}}_{t,i}^\textrm{phn}$ closer to positive samples $\bm{\hat{E}}_{a,j}^\textrm{phn}$, while the second term pushes the anchor $\bm{\hat{E}}_{a,i}^\textrm{phn}$ away from negative samples $\bm{\hat{E}}_{t,k}^\textrm{phn}$. 

The hyperparameters $\alpha$, $\beta$, and $\lambda$ control the embedding space boundaries and penalty levels for violations, which require manual tuning. 
To automate this, we adopt the AdaMS \cite{Jung-INTERSPEECH-AdaMS} framework. 
Unlike \cite{Jung-INTERSPEECH-AdaMS}, where hyperparameters are defined at the word level, we define them as learnable parameters specific to each phoneme class.

%highlighted in the blue area in Fig. \ref{fig:fig1}
\subsection{Utterance-Level Matching}
For utterance-level matching, highlighted in the blue area in Fig. \ref{fig:fig1}, we apply pooling operations to $\bm{E}^\textrm{phn}_a$ and $\bm{E}^\textrm{phn}_t$ to generate utterance-level embeddings, $\bm{E}^\textrm{utt}_a \in \mathbb{R}^{d}$ and $\bm{E}^\textrm{utt}_t \in \mathbb{R}^{d}$. 
Specifically, channel- and context-dependent statistics pooling (CCSP) \cite{Desplanques-arxiv} is used for $\bm{E}^\textrm{phn}_a$, while global average pooling is applied to $\bm{E}^\textrm{phn}_t$.
The utterance-level DML loss ($\mathcal{L}_\textrm{utt}$) is defined using RP loss.
The primary objective of RP Loss is to transfer structural information from TEs to AEs by preserving distance- and angle-wise relationships among TEs.
Building on our previous work \cite{RPL-Jung-INTERSPEECH}, we employ all three variants of RP loss: distance-wise RP ($\mathcal{L}_\textrm{RPL-D}$), angle-wise RP ($\mathcal{L}_\textrm{RPL-A}$), and prototypical RP ($\mathcal{L}_\textrm{RPL-P}$) losses.

In \cite{RPL-Jung-INTERSPEECH}, we demonstrated that incorporating a keyword classification loss ($\mathcal{L}_\textrm{key}$) as an auxiliary objective for the acoustic encoder significantly enhances open-vocabulary KWS performance.
In this work, we adopt the SphereFace2 framework, originally proposed for face recognition \cite{Wen-ICLR} and later adapted for speaker verification \cite{Bing-ICASSP}. 
Unlike conventional softmax normalization, which operates under a closed-set assumption, SphereFace2 loss utilizes multiple binary classifiers to train the model in a pair-wise manner, rather than performing multi-class classification.
Building on prior studies demonstrating that SphereFace2 outperforms multi-classification-based loss functions, we extend this framework to open-vocabulary KWS and show that SphereFace2 is also effective for this task.
To implement this, we convert the PyTorch-based official code provided by \cite{Bing-ICASSP}\footnote{\url{https://github.com/Hunterhuan/sphereface2_speaker_verification}} into TensorFlow.

% a weighted summation of the individual loss components, where the loss weight $\lambda_\textrm{phn}$ is set to 0.1:
The total loss function for embedding learning ($\mathcal{L}_\textrm{emb}$) is defined as follows (with $\lambda_\textrm{phn}=0.1$): 
\begin{equation}\label{eqn:eq3}
 \mathcal{L}_\textrm{emb} = \mathcal{L}_\textrm{utt} + \mathcal{L}_\textrm{key} + \mathcal{L}_\textrm{MM} + \lambda_\textrm{phn}\mathcal{L}_\textrm{phn}.
\end{equation}

\subsection{Modality Adversarial Learning}

The modality classifier $M$, with parameters $\theta_M$, consists of two fully connected layers with 256 hidden nodes. 
The classifier takes a batch of AEs ($\bm{\hat{E}}^\textrm{phn}_a$, $\bm{E}^\textrm{utt}_a$) and TEs ($\bm{\hat{E}}^\textrm{phn}_t$, $\bm{E}^\textrm{utt}_t$) as inputs and then predicts their modality labels (audio or text).
% We define two types of inter-modal adversarial loss ($\mathcal{L}_\textrm{adv}$) for phoneme-level ($\textrm{phn}$) and utterance-level ($\textrm{utt}$) embeddings, formulated as:
To encourage the encoders to generate modality-invariant representations, we define inter-modal adversarial losses at both the phoneme ($\textrm{phn}$) and utterance ($\textrm{utt}$) levels as follows:
% \begin{equation}\label{eqn:eq4}
%  \mathcal{L}^k_\textrm{adv} = -\frac{1}{N} \sum_{m \in {a,t}} \sum_{i=1}^{N} (o_{m,i}^k \log(M(E_{m,i}^k;\theta_M)), k \in \{\textrm{phn}, \textrm{utt}\},
% \end{equation}
\begin{align}\label{eqn:eq4}
\begin{split}
 \mathcal{L}^\textrm{phn}_\textrm{adv} = -\frac{1}{N^*} \sum_{m \in {a,t}} \sum_{i=1}^{N^*} (\bm{o}_{m,i}^\textrm{phn} \log(M(\hat{\bm{E}}_{m,i}^\textrm{phn};\theta_M)), \\
 \mathcal{L}^\textrm{utt}_\textrm{adv} = -\frac{1}{N} \sum_{m \in {a,t}} \sum_{i=1}^{N} (\bm{o}_{m,i}^\textrm{utt} \log(M(\bm{E}_{m,i}^\textrm{utt};\theta_M)),
\end{split}
\end{align}
where $\bm{o}_{m,i}^k$ is the ground-truth modality label of the input embedding vector for $k \in \{\textrm{phn}, \textrm{utt}\}$, expressed as a one-hot vector. 
Here, $M(\cdot;\theta_M)$ denotes the predicted modality probability of the input, and $\theta_M$ is shared for both $\mathcal{L}^\textrm{phn}_\textrm{adv}$ and $\mathcal{L}^\textrm{utt}_\textrm{adv}$. 
The total adversarial loss $\mathcal{L}_\textrm{adv}$ is the summation of $\mathcal{L}^\textrm{phn}_\textrm{adv}$ and $\mathcal{L}^\textrm{utt}_\textrm{adv}$.

The core of the proposed framework lies in the interaction between two key components: the multi-modal encoders (acoustic and text encoders) and the modality classifier, structured as a minimax game. 
The encoders aim to confuse the modality classifier by generating modality-invariant embeddings.
In contrast, the modality classifier attempts to distinguish embeddings based on their modality, thereby indirectly guiding the encoders to improve modality invariance.
This adversarial optimization improves cross-modal alignment, leading to more effective modality-invariant representation learning.

The training process jointly minimizes $\mathcal{L}_\textrm{emb}$ and $\mathcal{L}_\textrm{adv}$.
Following \cite{pmlr-v37-ganin15}, minimax optimization is implemented using a gradient reversal layer (GRL), which keeps the input unchanged during forward propagation and multiplies the gradient by -1 during back propagation.
%, effectively reversing the gradient flow.
By inserting a GRL before the modality classifier, the following two concurrent sub-processes are performed to obtain the updated parameters $\hat{\theta}_{\textrm{emb}}$ and $\hat{\theta}_{M}$: 
\begin{align}\label{eqn:eq5}
\begin{split}
 \hat{\theta}_{\textrm{emb}} = \argmin_{\theta_{\textrm{emb}}}(\mathcal{L}_\textrm{emb}(\theta_{\textrm{emb}})-\lambda_\textrm{adv}\mathcal{L}_\textrm{adv}(\hat{\theta}_{M})),\\
 \hat{\theta}_{M} = \argmax_{\theta_{M}}(\mathcal{L}_\textrm{emb}(\hat{\theta}_{\textrm{emb}})-\lambda_\textrm{adv}\mathcal{L}_\textrm{adv}(\theta_{M})),
\end{split}
\end{align}
where $\theta_\textrm{emb}$ denotes all trainable parameters for embedding learning except $\theta_{M}$ (i.e., encoders, cross-attention layer, CCSP layer, and keyword classifier). We set $\lambda_\textrm{adv}=0.1$.

\section{Experiments}
\subsection{Experimental Setup}

For training, we used the King-ASR-066 dataset \cite{Speechocean-DB}, following the configuration in \cite{RPL-Jung-INTERSPEECH}, including data augmentation.
This resulted in 4.6k hours of word-level speech segments, comprising 210k unique word classes.
For evaluation, we employed two benchmark datasets: Wall Street Journal (WSJ) \cite{Paul-WSN} and LibriPhrase \cite{Shin-INTERSPEECH}.
The evaluation setups followed \cite{RPL-Jung-INTERSPEECH} for WSJ and \cite{Shin-INTERSPEECH} for LibriPhrase.
The WSJ dataset contains 3.2k unique words with 18k word-level segments, augmented with synthetic Room Impulse Responses (RIRs) \cite{Ko-ICASSP} and MUSAN noises \cite{Snyder-arxiv} to simulate real-world conditions.
For the WSJ test set, we used 18k positive pairs and randomly selected 50 times more negative pairs, creating a highly imbalanced trial to better reflect real-world conditions.
To handle this imbalance, we evaluated model performance using Average Precision (AP) \cite{He-ICLR, Jung-INTERSPEECH, Hu-INTERSPEECH, Jung-ASRU}, which is particularly well-suited for imbalanced datasets.
For LibriPhrase, we followed the official trial setup and evaluation metrices, i.e., Equal-Error Rate (EER) and Area Under the ROC Curve (AUC).
For input acoustic features, we extracted 40-dimensional log Mel-filterbank coefficients with a frame length of 25 ms and a frame shift of 10 ms, applying mean normalization across each utterance.

All experiments were implemented using TensorFlow. 
Training was performed with the AdamW optimizer, starting with a $10^{-4}$ learning rate, which was halved every 20 epochs, and a weight decay of $10^{-5}$.
The model was trained for 100 epochs, taking approximately one day on two A100 GPUs.
Each mini-batch contained 500 utterances from 250 keywords, with two utterances per keyword.
For the AsyP loss in Eq. (\ref{eq2}), we set $\alpha = 0.01$, $\beta = 1.5$, and $\lambda = 0.01$. 
These values were later dynamically adjusted within the AdaMS framework.

For the acoustic encoder, we adopted ECAPA-TDNN \cite{Desplanques-arxiv}, widely used in prior studies \cite{Li-INTERSPEECH, RPL-Jung-INTERSPEECH, Jung-ICASSP}, with 256 filters and 1.8M parameters.
The text encoder consisted of a two-layer bi-LSTM with 256 hidden units. 
Input text was tokenized into G2P units \cite{ParkG2P}, then mapped to 256-dimensional embeddings via a trainable lookup table. The bi-LSTM processed these sequences, followed by global average pooling and a fully connected layer, producing a 256-dimensional TE.
The architectural design was optimized for the computational constraints of the target deployment environment.
During inference, only the two encoders and pooling layers were used, and the cosine similarity between $\bm{E}^\textrm{utt}_a$ and $\bm{E}^\textrm{utt}_t$ was computed for final matching.

\subsection{Comparison of Phoneme-Level Matching Methods}

\begin{table}[]
\renewcommand\thetable{1}
\caption{Comparison of phoneme-level matching losses. phn-level func. denotes the type of phoneme-level matching function.}
\vspace{-0.25cm}
\label{tab:tab1}
\scriptsize   
\centering
\begin{tabular}{cc|c|c}
\hline
\multicolumn{2}{c|}{Method}                            & \multirow{2}{*}{phn-level func.} & \multirow{2}{*}{AP (\%)} \\ \cline{1-2}
\multicolumn{1}{c|}{utt-level} & phn-level &         &       \\ \hline
\multicolumn{1}{c|}{RP loss}        & -         & -       & 75.10 \\ 
\multicolumn{1}{c|}{RP loss}        & Proxy-BD loss \cite{Yi-ICPR}  & MSP      & 83.72 \\ 
\multicolumn{1}{c|}{RP loss}        & Proxy-MS loss \cite{Wang-CVPR}  & ELSE    & 84.32 \\ 
\multicolumn{1}{c|}{RP loss}        & clat loss \cite{Kewei-arxiv}     & InfoNCE  & 83.59 \\ \hline\hline
\multicolumn{1}{c|}{\textbf{RP loss}} & \textbf{AsyP loss}       & ELSE+MSP                         & 84.42                    \\ 
\multicolumn{1}{c|}{\textbf{RP loss}} & \textbf{AsyP loss with AdaMS} & ELSE+MSP                         & \textbf{85.54}                    \\ \hline
\vspace{-0.4cm}
\end{tabular}
\end{table}

Table \ref{tab:tab1} compares our proposed phoneme-level matching method with existing methods on the WSJ dataset, including Proxy-BD loss \cite{Yi-ICPR}, Proxy-MS loss \cite{Wang-CVPR}, and clat loss \cite{Kewei-arxiv}. 
Proxy-BD and Proxy-MS losses are reformulated from the original BD and MS losses, as presented in \cite{Jung-INTERSPEECH}.
For a fair comparison, all methods utilize the same experimental setup, including datasets and model architectures, differing only in the choice of phoneme-level matching loss ($\mathcal{L}_\textrm{phn}$).
To isolate the impact of $\mathcal{L}_\textrm{phn}$, we exclude both $\mathcal{L}_\textrm{key}$ and $\mathcal{L}_\textrm{adv}$ from Eq. \ref{eqn:eq3} and Eq. \ref{eqn:eq4}.
As a baseline, we evaluate RP loss \cite{RPL-Jung-INTERSPEECH}, which applies only the utterance-level matching loss ($\mathcal{L}_\textrm{utt}$) without $\mathcal{L}_\textrm{phn}$.

The results indicate that all phoneme-level matching losses outperform the RP loss baseline, demonstrating the effectiveness of our cross-attention-based phoneme-level alignment.
In particular, our AsyP loss achieves the best performance among phoneme-level matching methods, with an 84.42\% AP. 
This highlights the importance of selecting an appropriate function for $\mathcal{L}_\textrm{phn}$, where using the ELSE function for the anchor-positive term and the MSP function for the anchor-negative term yields the best results, consistent with the trend observed in utterance-level DML loss \cite{Jung-INTERSPEECH}.
Furthermore, incorporating the AdaMS approach improves performance to 85.54\% AP, achieving a relative improvement of 8.40\% compared to the RP loss baseline.

\subsection{Performance of Modality Adversarial Learning}

\begin{table}[]
\renewcommand\thetable{2}
\caption{Performance of modality adversarial learning. DML indicates deep metric learning for utterance- or phoneme-level matching. MAL denotes modality adversarial learning.}
\vspace{-0.25cm}
\label{tab:tab2}
\scriptsize   
\centering
\begin{tabular}{l|cc|cc|c}
\hline
\multirow{2}{*}{Method}           & \multicolumn{2}{c|}{utt-level} & \multicolumn{2}{c|}{phn-level} & \multirow{2}{*}{AP (\%)} \\ \cline{2-5}
                              & \multicolumn{1}{c|}{DML} & MAL & \multicolumn{1}{c|}{DML} & MAL &                \\ \hline
\multirow{1}{*}{RP loss \cite{RPL-Jung-INTERSPEECH}}  & \multicolumn{1}{c|}{\checkmark}   & -   & \multicolumn{1}{c|}{-}   & -   &  75.10          \\  \hline
\multirow{4}{*}{\textbf{Proposed}} & \multicolumn{1}{c|}{\checkmark}   & \checkmark   & \multicolumn{1}{c|}{-}   & -   & 81.41                      \\   
                              & \multicolumn{1}{c|}{\checkmark}   & -   & \multicolumn{1}{c|}{\checkmark}   & -   & 85.54          \\  
                              & \multicolumn{1}{c|}{\checkmark}   & -   & \multicolumn{1}{c|}{\checkmark}   & \checkmark   & 86.12          \\  
                              & \multicolumn{1}{c|}{\checkmark}   & \checkmark   & \multicolumn{1}{c|}{\checkmark}   & \checkmark   & \textbf{86.23} \\ \hline
\multirow{2}{*}{CMCD \cite{Shin-INTERSPEECH}}  & \multicolumn{1}{c|}{-}   & -   & \multicolumn{1}{c|}{-}   & -   & 46.17          \\  
                              & \multicolumn{1}{c|}{-}   & -   & \multicolumn{1}{c|}{-}   & \checkmark   & 49.60               \\ \hline
\multirow{2}{*}{PhonMatchNet \cite{Lee-INTERSPEECH}} & \multicolumn{1}{c|}{-}   & -   & \multicolumn{1}{c|}{-}   & -   & 69.97          \\  
                              & \multicolumn{1}{c|}{-}   & -   & \multicolumn{1}{c|}{-}   & \checkmark   & 72.45              \\ \hline
\end{tabular}
\vspace{-0.3cm}
\end{table}

Table \ref{tab:tab2} evaluates the impact of Modality Adversarial Learning (MAL) on phoneme- and utterance-level matching using the WSJ dataset.
Comparing the first two rows, we observe that applying MAL only at the utterance level (without phoneme-level matching loss) boosts performance from 75.10\% AP to 81.41\% AP.
This configuration applies only $\mathcal{L}^\textrm{utt}_\textrm{adv}$ from Eq. (\ref{eqn:eq4}).
Here, the first row corresponds to the RP loss baseline from Table \ref{tab:tab1}.
Next, the third row represents the best-performing model from Table \ref{tab:tab1} (AsyP+AdaMS), which includes both utterance- and phoneme-level DML losses.
By further adding phoneme-level MAL, performance improves from 85.54\% to 86.12\% AP.
Finally, when MAL is applied at both utterance and phoneme levels, we achieve 86.23\% AP, marking a 14.82\% relative improvement over the RP loss baseline.

Beyond our proposed method, we also apply MAL to two baseline approaches, CMCD \cite{Shin-INTERSPEECH} and PhonMatchNet \cite{Lee-INTERSPEECH}, both of which use binary cross-entropy loss.
We reimplemented these baselines using the official code from \cite{Lee-INTERSPEECH}\footnote{\url{https://github.com/ncsoft/PhonMatchNet}}.
To ensure a fair comparison, all methods share the same experimental setup, including datasets and model architectures.
Since both CMCD and PhonMatchNet extract only phoneme-level embeddings without utterance-level embeddings, we apply only $\mathcal{L}^\textrm{phn}_\textrm{adv}$.
Notably, we observe that MAL enhances performance not only for our model but also for these baseline methods, further validating its effectiveness in reducing cross-modal discrepancies.

\subsection{Performance of Keyword Classification Loss}

\begin{table}[t]
\renewcommand\thetable{3}
\caption{Comparison of keyword classification losses.}
\vspace{-0.3cm}
\label{tab:tab3}
\scriptsize   
\centering
\begin{tabular}{c|c|c}
\hline
Method                                & $\mathcal{L}_\textrm{key}$               & AP (\%)        \\ \hline
\multirow{4}{*}{\textbf{ADML (ours)}} & -                    & 86.23          \\  
                                      & Triplet loss \cite{RPL-Jung-INTERSPEECH}         & 86.25          \\ 
                                      & AAM-Softmax loss \cite{Liu_2017_CVPR}     & 87.44          \\ 
                                      & \textbf{SphereFace2 (SF2)} & \textbf{90.24} \\ \hline
\end{tabular}
\vspace{-0.2cm}
\end{table}

Table \ref{tab:tab3} compares different keyword classification loss functions ($\mathcal{L}_\textrm{key}$) on the WSJ dataset.
``ADML (ours)'' refers to the best-performing model from Table \ref{tab:tab2}, incorporating DML and MAL at both the phoneme and utterance levels.
Applying triplet loss, previously used in \cite{RPL-Jung-INTERSPEECH}, yields only a marginal improvement, with the AP increasing slightly from 86.23\% to 86.25\%, which differs from the findings in \cite{RPL-Jung-INTERSPEECH}.
AAM-Softmax loss \cite{Liu_2017_CVPR}, widely used in classification tasks, performs significantly better, achieving 87.44\% AP.
Finally, SphereFace2 delivers the best performance, reaching 90.24\% AP, marking a 20.16\% relative improvement over the RP loss baseline.
This trend is consistent with previous findings in face recognition \cite{Wen-ICLR} and speaker verification \cite{Bing-ICASSP} tasks.

\begin{table}[]
\renewcommand\thetable{4}
\caption{Performance comparison on LibriPhrase.}
\vspace{-0.3cm}
\label{tab:tab4}
\scriptsize   
\centering
\begin{tabular}{c|cc|cc}
\hline
\multirow{2}{*}{Method} & \multicolumn{2}{c|}{EER (\%)}      & \multicolumn{2}{c}{AUC (\%)}      \\ \cline{2-5} 
                        & \multicolumn{1}{c|}{\textbf{LP\textsubscript{E}}} & \textbf{LP\textsubscript{H}} & \multicolumn{1}{c|}{\textbf{LP\textsubscript{E}}} & \textbf{LP\textsubscript{H}} \\ \hline
CMCD \cite{Shin-INTERSPEECH}              & \multicolumn{1}{c|}{4.20}  & 25.94 & \multicolumn{1}{c|}{99.12} & 81.14 \\ 
PhonMatchNet \cite{Lee-INTERSPEECH}        & \multicolumn{1}{c|}{2.33}  & 24.11 & \multicolumn{1}{c|}{99.70} & 83.39 \\ 
RP loss \cite{RPL-Jung-INTERSPEECH}            & \multicolumn{1}{c|}{1.54}  & 22.45 & \multicolumn{1}{c|}{99.84} & 85.31 \\ \hline\hline
\textbf{ADML+SF2 (ours)} & \multicolumn{1}{c|}{\textbf{1.33}} & \textbf{20.09} & \multicolumn{1}{c|}{\textbf{99.86}} & \textbf{88.71} \\ \hline
\end{tabular}
\vspace{-0.4cm}
\end{table}

\subsection{Performance Comparison on LibriPhrase}
We evaluate the performance on the LibriPhrase dataset, which consists of a hard set \textbf{LP\textsubscript{H}} and an easy set \textbf{LP\textsubscript{E}}.
% It is important to note that the models are trained on the King-ASR-066 dataset, not the LibriPhrase training set.
To evaluate generalization, all models are trained on King-ASR-066 and tested on LibriPhrase without access to its training set.
As in previous subsections, to ensure a fair comparison, all methods share the same experimental setup, including datasets and model architectures. 
The proposed method corresponds to the best model from Table \ref{tab:tab3}, referred to as ``ADML+SF2.'' 
We observe that our approach achieves the best performance on the LibriPhrase dataset, consistent with the results on the WSJ dataset.

\section{Conclusion}
In this work, we proposed Adversarial Deep Metric Learning (ADML), which integrates Modality Adversarial Learning (MAL) into the deep metric learning (DML) framework for open-vocabulary KWS.
To improve phoneme-level alignment, we introduced cross-attention-based phoneme-level matching with the Asymmetric-Proxy (AsyP) loss, enhanced by Adaptive Margins and Scaling (AdaMS). 
To address cross-modal heterogeneity between audio and text embeddings, we introduced MAL. 
Additionally, we applied SphereFace2-based keyword classification loss, further improving intra-modal discrimination.
Our experimental results on the WSJ and LibriPhrase datasets demonstrate that the proposed ADML outperforms existing methods. 
Future work will explore more advanced adversarial learning techniques to further improve MAL.

\bibliographystyle{IEEEtran}
\bibliography{mybib}

\end{document}